
\magnification\magstep1

\openup .5\jot

\input mssymb
\def\hbar{\mathchar '26\mkern -9muh}

\catcode`@=11
\def\eqaltxt#1{\displ@y \tabskip 0pt
  \halign to\displaywidth {%
    \rlap{$##$}\tabskip\centering
    &\hfil$\@lign\displaystyle{##}$\tabskip\z@skip
    &$\@lign\displaystyle{{}##}$\hfil\tabskip\centering
    &\llap{$\@lign##$}\tabskip\z@skip\crcr
    #1\crcr}}
\def\eqallft#1{\displ@y \tabskip 0pt
  \halign to\displaywidth {%
    $\@lign\displaystyle {##}$\tabskip\z@skip
    &$\@lign\displaystyle{{}##}$\hfil\crcr
    #1\crcr}}
\catcode`@=12 

\font\exmafy=msym10

\def\rrrm{{\hbox{\exmafy\char'122}}}

\def\half{{\textstyle {1 \over 2}}}

\def\pmb#1{\setbox0=\hbox{#1}  \kern-.025em\copy0\kern-\wd0
  \kern0.05em\copy0\kern-\wd0  \kern-.025em\raise.0433em\box0 }
\def\pmbh#1{\setbox0=\hbox{#1} \kern-.12em\copy0\kern-\wd0
            \kern.12em\copy0\kern-\wd0\box0}
\def\sqr#1#2{{\vcenter{\vbox{\hrule height.#2pt
      \hbox{\vrule width.#2pt height#1pt \kern#1pt
         \vrule width.#2pt}
      \hrule height.#2pt}}}}

\def\rchi{{\raise 2pt \hbox {$\chi$}}}
\def\rga{{\raise 2pt \hbox {$\gamma$}}}
\def\rg{{\raise 2 pt \hbox {$g$}}}

\def\susy{supersymmetry}
\def\({\left(}
\def\){\right)}
\def\<{\left\langle}
\def\>{\right\rangle}

\def\[{\left[}
\def\]{\right]}
\let\text=\hbox
\def\pt{\partial}
\def\eps{\epsilon}

\def\al{\alpha}

\def\om{\omega}
\def\ol{\overline}

\def\de{\delta}
\def\lam{\lambda}

\def\sig{\sigma}

\def\ti{\tilde}

\def\cH{{\cal H}}

\def\cD{{\cal D}}

\def\Lam{\Lambda}

\def\wti{\widetilde}
\def\Ga{\Gamma}

\hfuzz 6pt

\catcode`@=12 
\rightline {\bf DAMTP R94/24}
\rightline {\bf REVISED VERSION}
\centerline {\bf Quantization of a Locally Supersymmetric }
\centerline {\bf Friedmann Model with Supermatter\footnote{$\dagger$}{{\rm
Lecture
presented at the International School-Seminar
"Multidimensional Gravity and Cosmology", Yaroslavl,
 Russia, June 20 -- 26, 1994; to be published by World Scientific, ed. V.
Melnikov}} }
\vskip .1 true in
\centerline {A.D.Y. Cheng, P.D. D'Eath and
$\underline{{\rm P.R.L.V. Moniz}}$\footnote*{{\rm e-mail address:
prlvm10@amtp.cam.ac.uk}}}
\vskip .1 true in
\centerline {Department of Applied Mathematics and Theoretical Physics}
\centerline {University of Cambridge, Silver Street,Cambridge CB3 9EW, UK }
\vskip .2 true in
\centerline {\bf ABSTRACT}
\vskip .1 true in

{\sevenrm The  general theory of  N = 1
 supergravity with supermatter is studied using a
canonical approach. The supersymmetry and gauge constraint generators are
found.
The  framework is applied to the study of a Friedmann
minisuperspace model. We consider
a  Friedmann  k = + 1
geometry and  a family of spin-0 as well as spin-1 gauge
fields together with their odd (anti-commuting) spin-1/2 partners.
 The quantum supersymmetry constraints give rise to a set
of first-order coupled partial
differential equations for the components of the wave function.
As an intermediate stage in this  project, we put
 both the spin-1 field and its fermionic partner equal to zero.
 The  physical states of our simplified model
 correspond effectively to those of a mini-superspace
quantum cosmological model possessing N=4  local
supersymmetry coupled to   complex scalars with spin-1/2 partners.
The different supermatter models are given by specifying a
K\"ahler metric for the scalars;
the allowed quantum states then depend on the K\"ahler geometries.
For the cases of  spherically symmetric and flat K\"ahler geometries
we find  the general solution
for the quantum state with a very simple form. However,  although they allow a
Hartle-Hawking state, they do not allow a  wormhole state.}

\vfill
\magnification\magstep1
\centerline {PACS numbers: 04.60.+ $n$, 04.65.+ $e$, 98.80. $Hw$ }
\vskip 6 pt
\noindent

\noindent

\magnification\magstep1

\openup .5\jot

\input mssymb
\def\hbar{\mathchar '26\mkern -9muh}

\catcode`@=11
\def\eqaltxt#1{\displ@y \tabskip 0pt
  \halign to\displaywidth {%
    \rlap{$##$}\tabskip\centering
    &\hfil$\@lign\displaystyle{##}$\tabskip\z@skip
    &$\@lign\displaystyle{{}##}$\hfil\tabskip\centering
    &\llap{$\@lign##$}\tabskip\z@skip\crcr
    #1\crcr}}
\def\eqallft#1{\displ@y \tabskip 0pt
  \halign to\displaywidth {%
    $\@lign\displaystyle {##}$\tabskip\z@skip
    &$\@lign\displaystyle{{}##}$\hfil\crcr
    #1\crcr}}
\catcode`@=12 

\font\exmafy=msym10

\def\rrrm{{\hbox{\exmafy\char'122}}}

\def\half{{\textstyle {1 \over 2}}}

\def\pmb#1{\setbox0=\hbox{#1}  \kern-.025em\copy0\kern-\wd0
  \kern0.05em\copy0\kern-\wd0  \kern-.025em\raise.0433em\box0 }

\def\pmbh#1{\setbox0=\hbox{#1} \kern-.12em\copy0\kern-\wd0
            \kern.12em\copy0\kern-\wd0\box0}
\def\sqr#1#2{{\vcenter{\vbox{\hrule height.#2pt
      \hbox{\vrule width.#2pt height#1pt \kern#1pt
         \vrule width.#2pt}
      \hrule height.#2pt}}}}

\def\rchi{{\raise 2pt \hbox {$\chi$}}}
\def\rga{{\raise 2pt \hbox {$\gamma$}}}
\def\rrho{{\raise 2pt \hbox {$\rho$}}}

\def\({\left(}
\def\){\right)}
\def\<{\left\langle}
\def\>{\right\rangle}

\def\[{\left[}
\def\]{\right]}
\let\text=\hbox
\def\pt{\partial}
\def\eps{\epsilon}

\def\al{\alpha}

\def\om{\omega}
\def\ol{\overline}

\def\de{\delta}
\def\lam{\lambda}

\def\sig{\sigma}

\def\ti{\tilde}

\def\Lam{\Lambda}

\def\wti{\widetilde}
\def\Ga{\Gamma}

\def\cH{{\cal H}}


{\bf I. Introduction}

In the last ten years or so, the subjects of   supersymmetric
quantum gravity and cosmology have achieved a considerable number
of very interesting results and conclusions. Several approaches
may be found  in the literature, namely the triad ADM
canonical formulation [1--13], the $\sigma-$model
supersymmetric extension in quantum cosmology [14--16]
and another approach based on Ashtekar variables
 [17--20].
A detailed review on this  subject is currently in preparation [21].

The complete canonical quantization framework of N=1 (pure) supergravity was
presented in ref. [1]. It can be shown that it is sufficient,
in finding a physical state, to solve
the Lorentz and supersymmetry constraints of the theory  because the
algebra of constraints of the theory leads to anti-commutation relations
implying   that   a
physical wave functional $ \Psi $ will also obey
 the Hamiltonian constraints [1,22].

Using the triad ADM
canonical formulation, the  Bianchi-I model in $ N = 1 $ supergravity with no
cosmological constant ($ \Lam = 0 $) was considered in ref. [2] and
 the quantum states are
in the bosonic and filled fermionic sectors and are of the form
$\exp (-\half h^{-\half})$, where $ h = \det h_{i j} $ is the
determinant of the three-metric. In the case of Bianchi IX with $ \Lam = 0
$, there are
 two states, of the form $ \exp ( \pm I / \hbar ) $ where $ I $ is a
certain Euclidean action, one in the empty and one in the filled fermionic
sector [3,15].
When the usual choice of spinors constant in the standard basis is made
for the gravitino field, the bosonic state $ \exp ( - I / \hbar ) $ is the
wormhole state [3,24]. With a different choice, one obtains the
Hartle--Hawking state [23,25]. Similar states were found for $ N = 1 $
supergravity in the more general Bianchi models of class $ A $ [26].
[Supersymmetry (as well as other considerations) forbids mini-superspace
models of class $ B $.]  The extension of this analysis to the simple
case where  a cosmological constant is present in $ N = 1 $
supergravity is described in ref.[8--10]. It was  found
  by imposing the supersymmetry and Lorentz constraints that there
are then {\it no} physical states  in the models we have considered.
Regarding the  k=1 Friedmann-Robertson-Walker model, where the fermionic
degrees of freedom of the gravitino field
are very restricted,     we have found a bosonic quantum
physical state, namely  the Hartle-Hawking state for a De Sitter solution.
If one studies generic cosmological models using
 perturbation theory about the k=+1 Friedmann model, it seems that  the
gravitational and gravitino modes that are allowed to be excited in
 a supersymmetric Bianchi-IX model
contribute in such a way to
forbid any physical solutions of the quantum constraints.
This suggests that in a complete perturbation expansion we would have
to conclude that the
 full theory of N=1 supergravity with a non-zero cosmological constant
should have  {\it no} physical states.

  One would like to extend this
understanding to more general supergravity models involving lower-spin
fields.
 One possibility is to consider higher-$N$ gauged supergravity models
[34], but these are technically difficult in the approach used in [11] because
they contain a $ \Lam $-term which breaks chirality. Instead, we study here
the model of  $ N = 1 $ supergravity coupled to supermatter [27], and in
particular its supersymmetry constraints, especially in the case with zero
analytic potential $ P \( \Phi^I \) $.
Such a study was performed in ref.[13] for the case
 of  $ N = 1 $ supergravity coupled to supermatter [27].

The study of 1-dimensional mini-superspace models with local supersymmetry,
based on this, leads to further understanding of
quantum cosmology and gravity. Clearly, a  richer and more
 interesting class of minisuperspace models is given by
 coupling  supermatter to $ N = 1 $ supergravity
in 4 dimensions, and then
reducing the model to 1 dimension by making a suitable homogeneous Ansatz
[4--7]. In particular, from (1+3)  dimensional N=1 supergravity
a dimensional reduction allows  one to
obtain a (1+0)-dimensional theory with N=4 \susy .

In ref.[4--7]  an Ansatz for the gravitational and
 spin-3/2
fields was introduced in order to reduce
pure $ N = 1 $ supergravity in 4 dimensions to a locally supersymmetric
quantum cosmological model in 1 dimension, assuming a Friedmann $ k = + 1 $
geometry and homogeneity of the spin-3/2 field on the $ S^3 $ spatial
sections. The Hamiltonian structure of the resulting theory was found,
leading to the quantum constraint equations. The general solution to the
quantum constraints is very simple in this case, and the Hartle--Hawking
wave-function can be found. A more general model was also
studied, in which $ N = 1 $
supergravity is  coupled to locally supersymmetric matter, there taken to be a
massive complex scalar with spin-1/2 partner. In the massless case, the
general solution of the quantum constraints can be found as an integral
expression. Supergravity coupled to a massless complex scalar and
its spin-$\half$
partner also admits a ground quantum wormhole state [6]
 decribed by an integral expression. Other quantum wormhole states
can be found from it by simple differential operations.

Here we expand the study of mini-superspace quantum cosmological models
when  $ N = 1
$ supergravity is coupled to locally supersymmetric matter. We
consider the more
general
 supergravity  theory with
a  Friedmann $ k = + 1 $
geometry and  a family of spin-0 as well as spin-1 gauge
fields together with their odd (anti-commuting) spin-$\half$ partners.
The general such  theory is described in detail in ref. [27] (the
minisuperspace models with supermatter described  in
ref. [4--7] followed a four-dimensional model of Das et al [28]).
Our Ans\"atze for the  fields are   such as  to reduce
the   $ N = 1 $  supergravity plus supermatter in 4 dimensions [13,27]
to a locally supersymmetric N=4 FRW
quantum cosmological model in 1 dimension. Hence, we
  assume   a Friedmann $ k = + 1 $
geometry, and the other fields are chosen   as to respect
the homogeneity and isotropy of  the $ S^3 $ spatial
sections. The choice made for the spin-1 field is described in
ref. [29--31] and for the other fields the details
are given in ref. [4,5,7].
 The quantum supersymmetry constraints give rise to a set
of first-order coupled partial
differential equations for the components of the wave function.
As an intermediate stage in our research project, we put
 both the spin-1 field and its fermionic partner equal to zero.
 The  physical states of our simplified model
 correspond effectively to those of a mini-superspace
quantum cosmological model possessing N=4  local
supersymmetry coupled to   complex scalars with spin-1/2 partners.
The different supermatter models are given by specifying a
K\"ahler metric for the scalars;
the allowed quantum states then depend on the K\"ahler geometries.
For the cases of  spherically symmetric and flat K\"ahler geometries
we have  found the general solution
for the quantum state with a very simple form.
 However, these states  are somewhat
different from the ones presented in ref. [4--7]; although they allow a
Hartle-Hawking state, we cannot find a   wormhole state.

This paper is organized as follows. In section II
the more general theory of $ N = 1 $
 supergravity with supermatter is studied using a
canonical approach. The supersymmetry and gauge constraint generators are
also found.
In section III we specify our
Ans\"atze for the  the gravitational and
 spin-3/2
fields as well as for the supermatter fields and their fermionic
 partners.  The supersymmetry constraints are derived
from the reduced action in section IV. In section  V we
solve the quantum constraints and find a  general solution for the
 quantum state of the universe.  We also make  some comments on the
 issue of determining the operator ordering
in the constraints. A discussion and interpretation of our results
is presented in  section VI, together with a summary of our research
and indications of further possibilities.

\medbreak
\noindent
{\bf II.    Canonical Formulation  of $ N = 1 $
Supergravity with Supermatter               }

 The Lagrangian of the  more
general
gauged supergravity  theory coupled to a family of spin-0 as well as spin-1
fields together with their odd (anti-commuting) spin-$\half$ partners
is given in Eq. (25.12) of ref. [27]; it is too long to write out
here. It  depends on the tetrad
$ e^{A A'}_{~~~~\mu} $, where $ A, A' $ are two-component spinor indices
using the conventions of [1]
and $ \mu $ is a space-time index, the odd (anti-commuting) gravitino field $
\( \psi^A_{~~\mu}, \wti \psi^{A'}_{~~\mu} \) $, a vector field $ A^{(a)}_\mu $
labelled by an index $ (a) $, its odd spin-$\half $ partners $ \( \lam^{(a)}_A,
\wti \lam^{(a)}_{A'} \) $, a family of scalars $ \( \Phi^I, \Phi^{J^*}
\) $ and their odd spin-$\half$ partners $ \( \rchi^I_A, \wti \rchi^{J
^*}_{A'} \) $. The indices $ I, \ldots, J^*, \ldots $ are K\"ahler indices, and
there is a K\"ahler metric
$$ \rg_{I J^*} =   K_{I J^*} \eqno (2.1) $$
on the space of $ \( \Phi^I, \Phi^{J^*} \) $, where $ K_{I J^*} $ is a
shorthand
for $ \pt^2 K / \pt \Phi^I \pt \Phi^{J^*} $ with $ K $ the K\"ahler potential.
Each
index $ (a) $ corresponds to an independent Killing vector field of the
K\"ahler geometry. Such Killing vectors are holomorphic vector fields:
$$ \eqalignno {
X^{(b)} &= X^{I (b)} \( \Phi^J \)~{\pt \over \pt \Phi^I}~, \cr
X^{^* (b)} &= X^{I^* (b)} \( \Phi^{J^*} \)~{\pt \over \pt \Phi^{I^* }}~. &(2.2)
\cr
} $$
Killing's equation implies that there exist real scalar functions $ D^{(a)}
\( \Phi^I, \Phi^{I^*} \) $ known as Killing potentials, such that
$$ \eqalignno {
\rg_{I J^*} X^{J^* (a)} &= i~{\pt \over \pt \phi^I}~D^{(a)}~, \cr
\rg_{I J^*} X^{I (a)} &= - i~{\pt \over \pt \Phi^{J^*}}~D^{(a)}~. &(2.3) \cr
} $$

We shall consider instead  the Hamiltonian formulation of the theory [13].
 The
 Hamiltonian  has the form
$$ \eqalignno {
H = N \cH_\perp + N^i \cH_i + \psi^A_{~~0} S_A &+ \wti S_{A'} \wti
\psi^{A'}_{~~0} \cr
+ A^{(a)}_0 Q_{(a)} + M_{A B} J^{A B} &+ \wti M_{A' B'} \wti J^{A' B'}~, &(2.4)
\cr } $$
expected for a theory with the corresponding gauge invariances. Here $ N $
and $ N^i $ are the lapse function and shift vector [1], while $ \cH_\perp $
and $ \cH_i $ are the (modified) generators of deformations in the normal and
tangential directions. $ S_A $ and $ \wti S_{A'} $ are the local supersymmetry
generators, $ Q_{(a)} $ is the generator of gauge invariance, and $ J^{A B} $
and $ \wti J^{A' B'} $ are the generators of local
Lorentz rotations, while $ M_{A B} $ and $ \wti M_{A' B'} $ are Lagrange
multipliers giving the amount of Lorentz rotation applied per unit time.
Classically, the
constraints $ \cH_\perp, \cH_i, $ etc.~vanish, and the set of (first-class)
constraints forms an algebra.

 Quantum-mechanically, the constraints become
operators which annihilate physical states $ \Psi $:
$$ \eqalignno {
\cH_\perp \Psi &= 0~, \ \ \ \ \ \ \cH_i \Psi = 0~, \ \ \ \ \ \ S_A \Psi = 0~,
\ \ \ \ \ \ \ol S_{A'} \Psi = 0~, \cr
Q_{(a)} \Psi &= 0~, \ \ \ \ \ \ J^{A B} \Psi = 0~, \ \ \ \ \ \ \ol J^{A' B'}
\Psi =0~. &(2.5) \cr
} $$
Starting with the simplest of these, the $ J^{A B} $ and $ \ol J^{A' B'} $
quantum constraints imply that $ \Psi $ is constructed from Lorentz
invariants. The $ Q_{(a)} $ constraint, derived below, is of first order in
functional derivatives, and implies that the wave function $ \Psi $ is gauge
invariant. The $ S_A $ and $ \ol S_{A'} $ constraints will be derived and
discussed below. The $ \cH_\perp $ and $ \cH_i $ constraints can be defined
through the anti-commutator of $ S_A $ and $ \ol S_{A'} $, as in the case of $
N = 1 $ supergravity without matter fields [11]. Thus the remaining
constraints imply $ \cH_\perp \Psi = 0,~\cH_i \Psi = 0 $; if one could find a
solution of the remaining quantum constraints, the $ \cH_\perp $ and $ \cH_i
$ constraints would follow (with a certain choice of factor-ordering).

In the Hamiltonian decomposition, the variables are split into the spatial
components $ e^{A A'}_{~~~~i},~\psi^A_{~~i},~\wti \psi^{A'}_{~~i},~A^{(a)}_i,
 \lam^{(a)}_{A}, \wti \lam^{(a)}_{A'}, ~\Phi^I,~\Phi^{J^*},
\rchi^{I}_{A},\wti\rchi^{J^*}_{A'} $,
 which together with the bosonic momenta
are the basic dynamical variables of the theory, and the
Lagrange multipliers $ N,~N^i,~\psi^A_{~~0}$,$~\wti
\psi^{A'}_{~~0},$~$v^{(a)}_0,$
{}~$M_{A B}$,~$\wti M_{A' B'} $ of Eq.~(4), where $ N,
N^i $ are formed from the $ e^{A A'}_{~~~~0} $ and the $ e^{A A'}_{~~~~i} $
[6], and $ M_{A B}, \wti M_{A' B'} $ involve the zero components $ \om_{A B
0}, \wti \om _{A' B' 0} $ of the connection. One computes the canonical
momenta conjugate to the dynamical variables listed above in the usual way.
The constraint generators $ \cH_\perp, \cH_i $, etc.~are functions of the
basic dynamical variables. For the
gravitino and spin-$\half$ fields, the canonical momenta give second-class
constraints of the types described in [1,32,33]. These are eliminated when
Dirac
brackets are introduced [1,32,33] instead of the original Poisson brackets. In
particular, one obtains nontrivial Dirac brackets for $ p_{A A'}^{~~~~i} $,
the momentum conjugate to $ e^{A A'}_{~~~~i} $, for $ \psi^A_{~~i} $ and $
\wti \psi^{A'}_{~~i} $, for $ \lam^{(a)}_A $ and $ \wti \lam^{(a)}_{A'} $, for
$ \rchi^I_A $ and $ \wti \rchi^{J^*}_{A'} $, and for $ \pi_L,~\pi_{L^*} $,
the momenta conjugate to $ \Phi^L,~\Phi^{L^*} $. These can be made into simple
brackets by three steps.

First, the brackets involving $ p_{A A'}^{~~~~i},~\psi^A_{~~i} $ and $ \wti
\psi^{A'}_{~~i} $ can be simplified as in the case of pure $ N = 1 $
supergravity [1]. One redefines
$$ p_{A A'}^{~~~~i} \to \hat p_{A A'}^{~~~~i} = p_{A A'}^{~~~~i} - {1 \over
\sqrt 2} \eps^{i j k} \psi_{A j} \wti \psi_{A' k}~. \eqno (2.6) $$
This gives the Dirac brackets
$$
\[ \hat p_{A A'}^{~~~~i},~\psi^B_{~~j} \]^* =0~,~
\[ \hat p_{A A'}^{~~~~i},~\wti \psi^{B'}_{~~j} \]^*=0~,~
\[ \hat p_{A A'}^{~~~~i},~\hat p_{B B}^{~~~j} \]^* = ~{\rm independent \ of}
\ \psi^A_{~~i} \ {\rm and} \ \wti \psi^{A'}_{~~i}~, \eqno (2.7) $$

Next, one must deal with a complication caused by the dependence on the
scalars $ \Phi^I,~\Phi^{J^*} $ of the K\"ahler metric $ K_{I J^*} $ in the
second-class constraints. Defining $ \pi_{I A} $ to be the momentum conjugate
to $ \rchi^{I A} $, and $ \wti \pi_{I^* A'} $ to be the momentum
conjugate to $ \wti \rchi^{I^* A'} $, one has
$$
\pi_{I A} + {i e \over \sqrt 2}~K_{I J^*} n_{A A'} \wti \rchi^{J^* A'} = 0~, ~
\wti \pi_{J^* A'} + {i e \over \sqrt 2}~K_{I J^*} n_{A A'} \rchi^{I A} = 0~,
\eqno (2.8)  $$
where $ e = h^{1 \over 2} $, with $ h $ the determinant of the spatial metric
$ h_{i j} $. Here $ n^{A A'} $ is the spinor version of the unit
future-directed normal vector $ n^\mu $, obeying
$$ n_{A A'} n^{A A'} = 1~, \ \ \ \ \ \ n_{A A'} e^{A A'}_{~~~~i} = 0~. \eqno
(2.9) $$
The $ \Phi^K $ and $ \Phi^{K^*} $ dependence of $ K_{I J^*} $ is responsible
for
the unwanted Dirac brackets among $ \rchi^I_A,~\wti \rchi^{J^*}_{A},~\pi_L $
and $ \pi_{L^*} $. One cures this by using the square root of the K\"ahler
metric, $K^{1 \over 2}_{I J^*} $, obeying
$$ K^{1 \over 2}_{I J^*} \de^{K J^*} K^{1 \over 2}_{K L^*} = K_{I L^*}~. \eqno
(2.10) $$
This may be found by diagonalizing $ K_{I J^*} $ via a unitary transformation,
assuming that the eigenvalues are all positive. One needs to assume that
there is an ``identity metric'' $ \de^{K J^*} $ defined over the K\"ahler
manifold; this will be true if a positive-definite vielbein field can be
introduced. One then introduces the modified variables
$$
\hat \rchi_{I A} = e^{1 \over 2} K^{1 \over 2}_{I J^*} \de^{K J^*} \rchi_{K
A}~, ~
\hat {\wti \rchi}_{I^* A'} = e^{1 \over 2} K^{1 \over 2}_{J I^*} \de^{J
K^*} \wti \rchi_{K^* A'}~,\eqno  (2.11)
 $$
where the factor of $ e^{1 / 2} $ has been introduced for later use (in the
time gauge). Then the second-class constraints of Eq.(2.8) read
$$
\hat \pi_{I A} + {i \over \sqrt 2}~\de_{I J^*}~n_{A A'} \hat {\wti
\rchi}^{J^* A'} = 0~,
\hat {\wti \pi}_{I^* A'} + {i \over \sqrt 2}~\de_{I J^*}~n_{A A'} \hat
\rchi^{J A} = 0~. \eqno (2.12)  $$
The resulting Dirac brackets now give
$$
\[ \pi_L, \ \pi_M \]^* = 0~, ~
\[ \pi_L, \ \hat \rchi^A_{~~I} \]^* = 0~, \ {\rm etc.} \eqno (2.13)  $$

Finally, there are the brackets among $ \hat p_{A
A'}^{~~~~i},~\lam^{(a)}_A,~\wti \lam^{(a)}_{A'},~\hat \rchi^I_A $ and $
\hat {\wti \rchi}^{J^*}_{A'} $, which are just as in the case studied by
Nelson and Teitelboim [32]. These are dealt with by first defining
$$ \hat \lam^{(a)}_A = e^{1 \over 2} \lam^{(a)}_A~, \ \ \ \ \hat {\wti
\lam}^{(a)}_{A'} = e^{1 \over 2} \wti \lam^{(a)}_{A'}~. \eqno (2.14) $$
Then one goes to the time gauge, in which the tetrad component $ n^a $
of the normal vector $ n^\mu $ is henceforward restricted by
$$ n^a = \de^a_0~, \eqno (2.15) $$
or equivalently
$$ e^0_{~~i} = 0~. \eqno (2.16) $$
Thus the original Lorentz rotation freedom becomes replaced by that of
spatial rotations. In the time gauge, the geometry is described by the triad
$ e^\al_{~~i} (\al = 1, 2, 3) $, and the conjugate momentum is $ \hat
p_\al^{~~i} $. One has [33]
$$
\[ \hat p_\al^{~~i}, \ \hat p_\beta^{~~j} \]^{**} = 0~,~
\[ \hat p_\al^{~~i}, \ \hat \lam^{(a)}_A \]^{**} = 0~, ~
\[ \hat p_\al^{~~i}, \ \hat \rchi^I_{~~A} \]^{**} = 0~, \ {\rm etc.} \eqno
(2.17)
 $$
The remaining brackets are standard; the nonzero fermionic brackets are
$$ \eqalignno {
\[ \hat \lam^{(a)}~_A(x), \ \hat {\wti \lam}^{(b)}_{~~A'}(x) \]^{**} &=
\sqrt 2 i n_{A A'} \de^{(a) (b)} \de \( x, x' \)~, &(2.18) \cr
\[ \hat \rchi^I_{~~A} (x), \ \hat {\wti \rchi}^{J^*}_{~~A'} \( x' \) \]^{**}
&= \sqrt 2 i n_{A A'} \de^{I J^*} \de \( x, x' \)~, &(2.19) \cr
\[ \psi^A_{~~i} (x), \ \wti \psi^{A'}_{~~j} \( x' \) \]^{**} &= {1 \over
\sqrt 2} D^{A A'}_{~~~~i j} \de \( x, x' \)~, &(2.20) \cr } $$
where
$$ D^{A A'}_{~~~~i j} = - 2 i e^{- 1} e^{A B'}_{~~~~j} e_{B B' i} n^{B A'}~.
\eqno (2.21) $$

The supersymmetry constraint $ \wti S_{A'} $ is then found to be
$$ \eqalign {
\wti S_{A'} = &- \sqrt 2 i \hat \pi^{i j} e_{A A' i} \psi^A_{~~j} + \sqrt 2
\eps^{i j k} e_{A A' i}~^{3 s} \wti D_j \psi^A_{~~k} \cr
{}~&+ \( 2 \sqrt 2 \)^{- 1} e \wti \psi_{B'}^{~~[ j} \psi_B^{~~i ]} n^{B B'}
\psi^A_{~~j} e_{A A' i}
- \( \sqrt 2 \)^{- 1} e \wti \psi_{B'}^{~~[ i} \psi_B^{~~j ]} e^{B
B'}_{~~~~j} \psi^A_{~~i} n_{A A'} \cr
{}~&- {i \over \sqrt 2} \pi^{n (a)} e_{B A' n} \lam^{(a) B}
+ {1 \over 2 \sqrt 2} \eps^{i j  k} e_{B A' k} \lam^{(a) B} F^{(a)}_{i j}
+ {1 \over \sqrt 2} e g D^{(a)} n^A_{~~A'} \lam^{(a)}_{~~~A} \cr } $$
$$ + {1 \over \sqrt 2} \left [ \eqalign {
{}~&\pi_{J^*}
- {i e \over 2 \sqrt 2} n^{B B'} \lam^{(a)}_{~~~B} \wti \lam^{(a)}_{~~~B'}
K_{J^*} + {i \over \sqrt 2} e g_{L M^*} \Ga^{M^*}_{J^* N^*} n^{B B'} \wti
\rchi^{N^*}_{~~~B'} \rchi^L_{~~B} \cr
- &{i e \over 2 \sqrt 2} K_{J^*} g_{M M^*} n^{B B'} \wti \rchi^{M^*}_{~~~B'}
\rchi^M_{~~B}
- {1 \over 2 \sqrt 2} \eps^{i j k} K_{J^*} e^{B B'}_{~~~~j} \psi_{k B} \wti
\psi_{i B'} \cr
- &\sqrt 2 e g_{I J^*} \rchi^{I B} e_{B B'}^{~~~~m} n^{C B'} \psi_{m C} \cr }
\right ] \wti \rchi^{J^*}_{~~~A'} $$
$$ \eqalignno {
- &\sqrt 2 e g_{I J^*} \( \wti \cD_i \Phi^I \) \wti \rchi^{J^*}_{B'} n^{B B'}
e_{B A'}^{~~~~i} +
{i \over 2} g_{I J^*} \eps^{i j k} e_{A A' j} \psi^A_{~~i} \wti \rchi^{J^*
B'} e_{B B' k} \rchi^{I B} \cr
+ &{1 \over 4} e \psi_{A i} \( e_{B A'}^{~~~~i} n^{A C'} - e^{A C' i} n_{B A'}
\) g_{I J^*} \wti \rchi^{J^*}_{C'} \rchi^{I B} \cr
+ &{1 \over 4} e \psi_{A i} \( e_{B A'}^{~~~~i} n^{A C'} - e^{A C' i} n_{B
A'} \) \wti \rchi^{(a)}_{C'} \lam^{(a) B} \cr
- &e \exp ( K / 2 ) \[ 2 P n^A_{~~A'} e_{A B'}^{~~~~i} \wti
\psi^{B'}_{~~i} + i \( D_I P \) n_{A A'} \rchi^{I A} \]~, &(2.22) \cr
} $$
where $ \lam^{(a)}_A,~\wti \lam^{(a)}_{A'} $ and $ \rchi_{I A},~\wti
\rchi_{I^* A'} $ should be redefined as in Eqs.~(2.11),(2.14).

Here $ e_{A A' i} = \sig^\al_{A A'} e_{\al i} $, where $ \sig^\al_{A A'} (\al
= 1, 2, 3) $ are Infeld-van der Waerden symbols [1], and $ \hat \pi^{i j} = -
{1 \over 2} e^{\al ( i} \hat p_\al^{~~j )} $. Also [27]
$$ \eqalignno {
^{3 s} \wti \cD_j \psi^A_{~~k} = &\pt_j \psi^A_{~~k} + ~^{3 s} \om^A_{~~B j}
\psi^B_{~~k} \cr
+ &{1 \over 4} \( K_K \wti \cD_j \Phi^K - K_{K^*} \wti \cD_j \Phi^{K^*} \)
\psi^A_{~~k} \cr
+ &{1 \over 2} g A^{(a)}_j \( I m F^{(a)} \) \psi^A_{~~k}~, &(2.23) \cr } $$
where $ ^{3 s} \om_{A B j}, \ ^{3 s} \wti \om_{A' B' j} $ give the
torsion-free three-dimensional connection, and
$$ \wti \cD_i A^K = \pt_i A^K - g v^{(a)}_i X^{K (a)}~, \eqno (2.24) $$
with $ g $ the gauge coupling constant and $ X^{K (a)} $ the $ a $th Killing
vector field, as in Eq.(2.2). Further, the analytic functions $ F^{(a)} \(
\Phi^J
\) $ and $ F^{^* (a)} \( \Phi^{I^*} \) $ arise [27] from the transformation of
the
K\"ahler potential $ K $ under an isometry generated by the Killing vectors $
X^{(a)} $ and $ X^{^*(a)} $:
$$ \de K = \( \eps^{(a)} X^{(a)} + \eps^{^*(a)} X^{^* (a)} \) K ~. \eqno (2.25)
$$
One obtains
$$ \de K = \eps^{(a)} F^{(a)} - \eps^{^* (a)} F^{^*(a)} - i \( \eps^{(a)} -
\eps^{^*(a)} \) D^{(a)}~, \eqno (2.26) $$
where $ D^{(a)} $ is the Killing potential of Eq.(2.3). Also, in Eq.(2.22),
$\pi^{n (a)} $ is the momentum conjugate to $ A^{(a)}_n ,~K_{J^*} $ denotes
$ \pt K / \pt \Phi^{J^*},~\Ga^{M^*}_{J^* N^*} $ denotes the starred Christoffel
symbols [27] of the K\"ahler geometry, and $ P = P \( \Phi^I \) $ gives the
potential of the theory.

The gauge generator $ Q^{(a)} $ is given classically by
$$ \eqalignno {
Q^{(a)} &= - \pt_n \pi^{n (a)} - g f^{a b c} \pi^{n (b)} A^{(c)}_n \cr
&+ g \( \pi_I X^{I (a)} + \pi_{I^*} X^{I^* (a)} \) \cr
&+ \sqrt 2 i e g K_{M I^*} n^{A A'} X^{J^* (a)} \Ga^{I^*}_{J^* N^*} \wti
\rchi^{N^*}_{A'} \rchi^M_A \cr
&- \sqrt 2 i e g n^{A A'} \wti \lam^{(b)}_{A'} \[ f^{a b c} \lam^{(c)}_A +
{1 \over 2} i \( I m F^{(a)} \) \lam^{(b)}_A \] \cr
&+ \sqrt 2 i e g n^{A A'} K_{I J^*} \wti \rchi^{J^*}_{A'} \[ {\pt X^{I (a)}
\over \pt A^J} \rchi^J_A + {1 \over 2} i \( I m F^{(a)} \) \rchi^I_A \] \cr
&- {i \over \sqrt 2} g \( I m F^{(a)} \) \eps^{i j k} \wti \psi_{i A'} e^{A
A'}_{~~~~j} \psi_{A k}~, &(2.27) \cr } $$
where $ f^{a b c} $ are the structure constants of the isometry group.

One can proceed to a quantum description by studying (for example)
Grassmann-algebra-valued wave functions of the form
$ \Psi \( e^\al_{~~i},  \psi^A_{~~i},  A^{(a)}_{~~i},  \hat \lam^{(a)}_A,
  n_A^{~~A'} \hat {\wti \rchi}^{J^*}_{A'}, \Phi^I, \Phi^{J^*} \) $.
The choice of $ n_A^{~~A'} \hat {\wti \rchi}^{J^*}_{A'} $ rather than $ \hat
\rchi^J_A $ is designed so that the quantum constraint $ \ol S_{A'} $ should
be of first order in momenta. The momenta are represented by
$$ \eqalignno {
\hat p_\al^{~~i} &\to - i \hbar \de / \de e^\al_{~~i}~,~
\pi^{n (a)} \to - i \hbar \de / \de A^{(a)}_n~, ~
\pi_I \to - i \hbar \de / \de \Phi^I~,
\pi_{I^*} \to - i \hbar \de / \de \Phi^{I^*}~, &(2.28) \cr
\ol \psi^{A'}_{~~i} &\to  {1 \over \sqrt 2} i \hbar D^{A A'}_{~~~~j i} \de /
\de \psi^A_{~~j}~, ~
\hat {\ol \lam}^{(a) A'} \to - \sqrt 2 n^{A A'} \de  / \de \hat \lam^{(a)}~,
\hat \rchi^{I A} \to - \sqrt 2 n^{A A'} \de^{I J^*} \de / \de \hat {\wti
\rchi}^{J^* A'}~. &(2.29) \cr } $$

Quantum-mechanically, we order each term cubic in fermions in $ \ol S_{A'} $
(using anti-commutation) such that one ``momentum'' fermionic variable is on
the right, and two ``coordinate'' fermionic variables are on the left. The
ordering of the quantum constraint $ S_A $ is defined by taking the hermitian
adjoint with respect to the natural inner product [1,35]. Then the terms in $
S_A $ cubic in fermions have two ``momenta'' on the right and one
``coordinate'' on the left.

\medbreak
\noindent
{\bf III. Ans\"atze for the Fields and Dimensional Reduction}

In order to learn more about
quantum cosmology with local supersymmetry we would like to
study
 certain types of simple mini-superspace models.  Among the simplest
non-trivial mini-superspace models (in which the gravitational and matter
variables have been reduced to a finite number of degrees of freedom) are
those based on Friedmann universes with $ S^3 $ spatial sections,
which are the spatial
 orbits of $G=SO(4)$ -- the group of homogeneity and
isotropy.
Consistent with this assumption we choose the geometry
to be that of a $ k = + 1 $ Friedmann model. The tetrad of the
four-dimensional theory is taken to be:
$$ e_{a\mu} = \pmatrix {
N (\tau) &0 \cr
N^i a (\tau) E_{\hat a i} &a (\tau) E_{\hat a i} \cr }~,
\ \ \ e^{a \mu} = \pmatrix {
N (\tau)^{-1} &-N^i N (\tau)^{-1} \cr
0 &a (\tau)^{-1} E^{\hat a i} \cr }~ \eqno (3.1)$$
where $ \hat a $ and $ i $ run from 1 to 3. The shift vector $ N^i $ is
assumed to take the form $ N^i = - N^{AA'} (\tau) e_{AA'}^{~~~~i} $, where $
N^{AA'} n_{AA'} = 0 $.

$ E_{\hat a i} $ is a basis of left-invariant 1-forms on the unit $ S^3 $
with volume $ \sig^2 = 2 \pi^2 $. The spatial tetrad $ e^{AA'}_{~~~~i} $
satisfies the relation
$$ \pt_i e^{AA'}_{~~~~j} - \pt_j e^{AA'}_{~~~~i} = 2 a^2 e_{ijk} e^{AA'k}
\eqno (3.2) $$
as a consequence of the group structure of SO(3), the isotropy
(sub)group.

This Ansatz reduces the number of degrees of freedom provided by $ e_{AA'
\mu} $. If supersymmetry invariance is to be retained, then we need an
Ansatz for $ \psi^A_{~~\mu} $ and $ \ti \psi^{A'}_{~~\mu} $ which reduces the
number of fermionic degrees of freedom, so that there is equality between the
number of bosonic and fermionic degrees of freedom. One is naturally led to
take $
\psi^A_{~~0} $ and $ \ti \psi^{A'}_{~~0} $ to be functions of time only. In
the four-dimensional Hamiltonian theory, $ \psi^A_{~~0} $ and $ \ti
\psi^{A'}_{~~0} $ are Lagrange multipliers which may be freely specified. For
this reason we do not allow $ \psi^A_{~~0} $ and
 $ \ti \psi^{A'}_{~~0} $ to depend on $ \psi^A_{~~i} $ or
$ \ti \psi^{A'}_{~~i} $ in our Ansatz. We
further take
$$
\psi^A_{~~i} = e^{AA'}_{~~~~i} \ti \psi_{A'}~, ~
\ti \psi^{A'}_{~~i} = e^{AA'}_{~~~~i} \psi_A~, \eqno (3.3)  $$
where we introduce the new spinors $ \psi_A $ and $ \ti \psi_{A'} $ which
are functions of time only. [It is possible to justify the Ansatz (3.3) by
requiring that the form (3.1) of the tetrad be preserved under suitable
homogeneous supersymmetry transformations [4,5,7].]
Moreover, it turns out that
the constraints obeyed by classical solutions of the 1-dimensional theory
lead to a 4-dimensional energy-momentum tensor which is isotropic, consistent
with the assumption of a Friedmann geometry.

It is important to remark that
the Ansatz for $ \psi^A_{~~i} $ is preserved under a combination of
 a non-zero (spatially homogeneous)
supersymmetry transformation and
possible local Lorentz and coordinate transformations [4,5,7]
if we impose the additional
constraint
$$ \psi^B \ti \psi^{B'} e_{BB'i} = 0~. \eqno (3.4) $$
Eq. (3.4) implies that $ \psi^B \ti \psi^{B'} \propto n^{BB'} $,
and so we can write (2.8) in the two equivalent forms:
$$ \eqalignno {
J_{AB} = \psi_{(A} \ti \psi^{B'} n_{B)B'} &= 0~, &(3.5) \cr
\ti J_{A'B'} = \ti \psi_{(A'} \psi^B n_{BB')} &= 0~.
&(3.6) \cr
} $$
 The
constraint $ J_{AB} = 0 $ has a natural interpretation as the reduced form of
the Lorentz rotation constraint arising in the full theory [1].
By requiring that the constraint $ J_{AB} = 0 $ be preserved under the same
combination of transformations as used above, one finds equations
 which
are satisfied provided the supersymmetry constraints $ S_A = 0,~\ti S_{A'} =
0 $ (see  below) hold.
By further requiring that the supersymmetry
constraints be preserved, one finds additionally that the Hamiltonian
constraint $ {\cal H} = 0 $  should hold.
When matter fields are taken into account (see next paragraphs) the
generalisation of the $ J_{AB} $
constraint  is :
$$ J_{AB} = \psi_{(A} \ti \psi_{B)} - \rchi_{(A} \ti \rchi_{B)}
- \lam^{(a)}_{(A} \ti \lam^{(a)}_{B)} =  0~.
\eqno (3.7) $$
One can justify this by observing either that it arises from the
corresponding constraint of the full theory, or that its quantum version
describes the invariance of the wavefunction under Lorentz transformations.

Now, let us address the supermatter fields. First, we choose
for the gauge group of our model the  group  $\hat{G} = SU(2)$. In
this case we have that [27]
$$ D^{(1)}= \half \left({{\phi + \bar \phi}\over {1 + \phi\bar\phi}}\right),
{}~D^{(2)}= - {i \over 2} \left({{\phi - \bar \phi}\over {1 +
\phi\bar\phi}}\right),
{}~D^{(3)}=~- \half \left({{1 - \phi\bar\phi}\over {1 + \phi\bar\phi}}\right)
,\eqno(3,8)$$
and
$$ K = \ln(1 + \phi \bar \phi)~,\eqno(3.9)$$
and thus
$$ \rg_{\phi \bar\phi}= { 1 \over (1 + \phi \bar \phi)^{2} }
{}~,\rg^{\phi \bar\phi}=~(1 + \phi \bar \phi)^{2} ~.\eqno(3.10)$$
The Levi-Civita connections of the  $S^{2}$ K\"ahler manifold are
$$ \Gamma^{\phi}_{\phi \phi} = g^{\phi \bar \phi} { {\pt g_{\phi \bar \phi} }
\over {\pt\phi}} = { -2 \bar \phi \over (1 + \phi \bar \phi)}  ~\eqno(3.11)$$
and its complex conjugate. The rest of the components are zero.
The scalar super-multiplet, consisting of a complex massive scalar
field $ \phi $ and massive spin-1/2 field $ \rchi, \ti \rchi$
are chosen to be
spatially homogeneous, depending only on time. The
odd spin-$\half $ partner $ \(
\lam^{(a)}, \wti \lam^{(a)}\) $, $a=1,2,3$, are chosen to
depend only on time as well. As far as the
vector field $ A^{(a)}_\mu $
is concerned we adopt here the ansatz formulated in ref.
[29,30,31]. More specifically, since are the physical
observables to be $SO(4)$-invariant, the
fields with gauge degrees of freedom may
transform under $SO(4)$ if these
 transformations can be compensated by a gauge
transformation. This is so since the physical observables
are gauge invariant quantities. Fortunately
there is a large class of fields satisfying the above
conditions. These are the so-called {\it $SO(4)$--symmetric
fields}, i.e. fields which are invariant up to a gauge
transformation. According to group theory considerations
[29,30,31] the  {it $SO(4)$--symmetric}
spin-1 field is taken to be
$$
{\bf A}_{\mu}(t)~\omega^{\mu}  =
\left(
{{f(t)}\over {4}}
\varepsilon_{acb}{\cal T}^{(3)}_{ab}\right)\omega^c
{}~,\eqno(3.12)$$
where $\{\omega^{\mu}\}$ represents the moving coframe
$\{\omega^{\mu}\} = \{ dt,\omega^b\}~,~(b=1,2,3)~$,  of
one-forms, invariant under the left action of $SU(2)$ and ${\cal T}^{(3)}_{ab}$
are the generators of the $SU(2)$ gauge group.
The idea behind this Ansatz for a non-Abelian
spin-1 field is to define a homorphism of the
isotropy group $SO(3)$ to the gauge group.
This homomorphism defines the gauge transformation
which, for the symmetric fields,
 compensates the action of a given $SO(3)$ rotation.
Hence, the above form
for the gauge field  where the $A_0$ component is taken
to be identically zero.


If one assumes   that the dynamics of the most
general $N=1$ supergravity theory coupled to supermatter
is as given in Eq.(25.12) of [27] than,  by imposing the
above mentioned symmetry conditions, we obtain a
one-dimensional (mechanical) model depending only on $t$.
The resulting
one-dimensional  model will have some
symmetries remaining from the symmetries of the
four-dimensional theory. In particular the invariance
under general coordinate transformations in four
dimensions leads
to an invariance under arbitrary
time--reparametrizations. However, due to our choice
of  $SO(4)$--symmetry conditions on the spin-1 field,
none of the local internal
(i.e. gauge) symmetries will survive: all the available
  gauge transformations are required to
cancel out the action of a given $SO(3)$ rotation.
Thus, we will not have in our FRW case a gauge
constraint $Q^{(a)}=0$ [29--31].
However, in the case of larger gauge group
some of the
 gauge symmetries will survive, giving rise, in the
one-dimensional model, to local internal symmetries with
a reduced gauge group.
Therefore,   a gauge
constraint can be expected to play an important role
in such a case [13,29--31] and a study of such a model
would be  interesting.

In the next section
we will study  associated FRW
cosmological model. From the one-dimensional
 effective action we will derive the
supersymmetry constraints of our theory.

\medbreak
\medbreak
\noindent
{\bf IV.  Supersymmetry Constraints
in the  One-Dimensional Theory }

Using the Ans\"atze described in the previous section, the action of the full
theory
(Eq. (25.12) in ref. [27]) is reduced to one with a
 finite number of degrees of freedom.
Starting from the action so  obtained,
 we  study the Hamiltonian formulation of this model.
As discussed above, the Hamiltonian of any supersymmetric model has the form
(2.4).
 The procedure to find the expressions of $S_{A}$ and $\bar S_{A'}$ is very
simple.
 First, we have to calculate the conjugate momenta of the
 dynamical variables and then evaluate the usual expression:
$$ H(p_{i}, q^{i}) = p_{i} \dot{q}^{i} - L ~. \eqno (4.1)$$
Afterwards, we  read out the coefficients of $\psi_{0}^{~A}$    and
 $\bar \psi_{0}^{~A'}$   from this expression in order to get the $S_{A}$ and
$\bar S_{A'}$
constraints, respectively.

The contributions from the spin-0 field $\phi$ to the $\bar S_{A'}$ constraint
are seen to be
 $$ {1\over \sqrt{2}} \bar \rchi_{A'} \left[ \pi_{\bar \phi} + {i \sigma^{2}
a^{3} \over 2 \sqrt{2}}
{\phi \over (1 + \phi \bar \phi)} n_{BB'} \bar \lambda^{(a)B'} \lambda^{(a)B} -
{5i \over 2 \sqrt{2}}
{\sigma^{2}a^{3} \phi \over (1 + \phi \bar \phi)^{3}} n_{BB'} \bar \rchi^{B'}
\rchi^{B} \right.$$
$$ \left. - {3i \over 2 \sqrt{2}} { \sigma^{2} a^{3} \phi \over (1 + \phi \bar
\phi)} n_{BB'} \psi^{B} \bar \psi^{B'} - {3 \over \sqrt{2}}
{\sigma^{2}a^{3} \over (1 + \phi \bar \phi)^{2}} n_{BB'} \rchi^{B} \bar
\psi^{B'} \right]
+ {\sigma^{2}a^{2}gf \over \sqrt{2} (1 + \phi \bar \phi)^{2}} \sigma
^{a}_{~AA'}n^{AB'} \bar \rchi_{B'} X^{a}
{}~. \eqno (4.2) $$
The contributions to the $\bar S_{A'}$ constraint  from the spin-1 field are
$$ -i {\sqrt{2} \over 3} \pi_{f} \sigma^{a}_{~BA'} \lambda^{(a)B}
 + { \sigma^{2}a^{3} \over 6} \sigma^{a}_{~BA'}\lambda^{(a)B} n_{CB'}
(\sigma^{bCC'} \bar \psi_{C'} \bar \lambda^{(b)B'} +
\sigma^{bAB'} \psi_{A} \lambda^{(b)C})$$
$$+ {1 \over 8 \sqrt{2}}\sigma^2 a^4 \sigma^{(a)C}_{~~A'} [1 - (f - 1)^2]
\lambda^{(a)}_{~~C}$$
$$ + {1 \over 2} \sigma^{2}a^{3} \lambda^{(a)A}(-n_{AB'} \bar \psi_{A'} \bar
\lambda^{(a)B'} + n_{BA'} \psi_{A} \lambda^{(a)B} - {1 \over 2} n_{AA'}
\psi_{B} \lambda^{(a)B}
 + {1 \over 2} n_{AA'} \bar \psi_{B'}
\bar \lambda^{(a)B'})~. \eqno (4.3) $$
The contributions from the spin-2 field and spin-3/2 field to $\bar S_{A'}$
constraint are
 $$ {i \over 2 \sqrt{2}}a \pi_{a} \bar \psi_{A'} -{3 \over \sqrt{2}}
\sigma^{2}a^{2} \bar \psi_{A'}
+ {3 \over 8} \sigma^{2} a^{3} n^{B}_{~A'} \bar \psi^{B'} \psi_{B} \bar
\psi_{B'} ~. \eqno (4.4)$$
The following terms are  also present in the  $\bar S_{A'}$   supersymmetry
constraint:
$$ -{1 \over \sqrt{2}} \sigma^{2}a^{3} g D^{a} n_{AA'} \lambda^{(a)A}$$
$$ + {\sigma^{2}a^{3} \over (1 + \phi \bar \phi)^{2}}(- n_{BA'} \bar \psi_{B'}
+ {1 \over 2} n_{BB'} \bar \psi_{A'}) \rchi^{B} \bar \rchi^{B'}$$ $$- {1 \over
4} \sigma^{2}a^{3}(n_{AB'} \lambda^{(a)A} \bar \lambda^{(a)}_{~~A'} \bar
\psi^{B'} + n_{AA'} \lambda^{(a)A} \bar \lambda^{(a)}_{~~B'} \bar \psi^{B'})$$
$$  - {1 \over 4 (1 + \phi \bar \phi)^{2}} \sigma^{2}a^{3} (n_{AB'} \rchi^{A}
\bar \rchi_{A'} \psi^{B'} + n_{AA'}
\rchi^{A} \bar \rchi_{B'} \psi^{B'})~. \eqno (4.5)$$
The supersymmetry constraint $\bar S_{A'}$ is then
 the sum of the above expressions. The supersymmetry constraint $S_{A}$ is just
the
 complex conjugate of $\bar S_{A'}$.

\medbreak
\noindent
{\bf IV. Solutions of the Supersymmetry Constraints.}

In this section we will solve explicitely the corresponding quantum
supersymmetry constraints. As an intermediate stage in our research project, we
put
 both the spin-1 field and its fermionic partner equal to zero.
 The  physical states of our simplified model
 correspond effectively to those of a mini-superspace
quantum cosmological model possessing N=4  local
supersymmetry coupled to   complex scalars with spin-1/2 partners.
The different supermatter models are given by specifying a
K\"ahler metric for the scalars;
the allowed quantum states then depend on the K\"ahler geometries.
For the cases of  spherically symmetric and flat K\"ahler geometries
we have  found the general solution
for the quantum state with a very simple form.
 However, these states  are somewhat
different from the ones presented in ref. [4--7]; although they allow a
Hartle-Hawking state, we cannot find a   wormhole state.

 First we need to redefine the  $ \rchi_{A} $ field and $ \psi_{A} $ field in
order to simplify the Dirac
brackets [13,23], following some of the steps described in section II:

$$ \hat \rchi_{A} = {\sigma a^{3 \over 2} \over 2^{1 \over 4} (1 + \phi \bar
\phi)} \rchi_{A}~, ~  \bar \rchi_{A'} = {\sigma a^{3 \over 2} \over 2^{1 \over
4} (1 + \phi \bar \phi)} \bar \rchi_{A'} ~. \eqno(5.1) $$
The conjugate momenta become
$$ \pi_{\hat \rchi_{A}} = -i n_{AA'} \hat \rchi^{A'}~, ~
 \pi_{\hat \rchi_{A'}} = -i n_{AA'} \hat \rchi^{A} ~.\eqno(5.2) $$
This pair form a set of   second class constraints. Consequently, the Dirac
bracket becomes

$$ [ \hat \rchi_{A} , \hat \rchi_{A'} ]^{*}_{+} = -i n_{AA'}~. \eqno(5.3) $$
Similarly for the $ \psi_{A}$  field,

$$ \hat \psi_{A} = {\sqrt{3} \over 2^{1 \over 4}} \sigma a^{3 \over 2}
\psi_{A}~, ~
\hat \psi_{A'} = {\sqrt{3} \over 2^{1 \over 4}} \sigma a^{3 \over 2} \bar
\psi_{A'} ~,\eqno(5.4) $$
where the conjugate momenta are

$$ \pi_{\hat \psi_{A}} = in_{AA'} \hat \rchi^{A'} ~,~
\pi_{\hat \psi_{A'}} = in_{AA'} \hat \rchi^{A}~. \eqno(5.5)$$
The Dirac bracket is then

$$ [\hat \psi_{A} , \hat \psi_{A'}]^{*}_{+} = in_{AA'}~. \eqno(5.6) $$
Furthermore,
$$ [a , \pi_{a}]^{*} = 1~, ~ [\phi, \pi_{\phi}]^{*} = 1~,
 ~[\bar \phi, \pi_{\bar \phi}]^{*} = 1~, \eqno(5.7)$$
and the rest of the brackets are zero.

After substituting  the redefined  fields in the  constraints,
 we drop the hat over the new variables. The supersymmetry constraints become

$$ \bar S_{A'} = {1 \over \sqrt{2}} (1 + \phi \bar \phi) \bar \rchi_{A'}
\pi_{\bar \phi} + {i \over 2 \sqrt{6}} a \pi_{a} \bar \psi_{A'} $$
$$ - \sqrt{3 \over 2} \sigma^{2}a^{2} \bar \psi_{A'} - {5i \over 2 \sqrt{2}}
\phi n_{BB'} \bar \rchi^{B'} \rchi^{B} \bar \rchi_{A'} $$
$$-{1 \over 4 \sqrt{6}} n_{BA'} \bar \psi^{B'} \psi^{B} - {i \over 2 \sqrt{2}}
\phi n_{BB'} \psi^{B} \bar \psi^{B'} \rchi_{A'} $$
$$ -{5 \over 2 \sqrt{6}} n_{BB'} \rchi^{B} \bar \psi^{B'} \bar \rchi_{A'} -
{\sqrt{3} \over 2 \sqrt{2}} n_{AA'} \bar \psi_{B'} \rchi_{A} \bar \rchi^{B'} $$
$$ + {1 \over \sqrt{6}} n_{BB'} \rchi^{B} \bar \rchi^{B'} \bar \psi_{A'}
\eqno(5.8) $$
together with its complex conjugate.

It is simpler to describe the theory using only (say) unprimed spinors, and, to
this end, we define
$$ \bar \psi_{A} = 2 n_{A}^{~B'} \bar \psi_{B'}~, ~
 \bar \rchi_{A} = 2 n_{A}^{~B'} \bar \rchi_{B'} ~,\eqno(5.9) $$
with which the new Dirac brackets are
$$ [\rchi_{A}, \bar \rchi_{B}]^{*}_{+} = -i \epsilon_{AB}~, ~
 [\psi_{A}, \bar \psi_{B}]^{*}_{+} = -i \epsilon_{AB} ~.\eqno(5.10) $$
The rest of the brackets remain unchanged. Using these new variables, the
supersymmetry constraints are
$$ S_{A} = {1 \over \sqrt{2}} (1 + \phi \bar \phi) \rchi_{A} \pi_{\phi} - {i
\over 2 \sqrt{6}} a \pi_{a} \psi_{A} $$
$$ - \sqrt{3 \over 2} \sigma^{2}a^{2} \psi_{A} - {5i \over 4 \sqrt{2}} \bar
\phi \rchi_{A} \bar \rchi_{B} \rchi^{B} $$
$$+{1 \over 8 \sqrt{6}} \psi_{B} \bar \psi_{A}  \psi^{B} - {i \over 4 \sqrt{2}}
\bar \phi \rchi_{A} \psi^{B} \bar \psi_{B} $$
$$ +{5 \over 4 \sqrt{6}} \rchi_{A} \psi^{B} \bar \rchi_{B} + {\sqrt{3} \over 4
\sqrt{2}} \rchi^{B} \bar \rchi_{A} \psi_{B} $$
$$ - {1 \over 2 \sqrt{6}} \psi_{A} \rchi^{B} \bar \rchi_{B}  \eqno(5.11)$$
and

$$ \bar S_{A} = {1 \over \sqrt{2}} (1 + \phi \bar \phi) \bar \rchi_{A}
\pi_{\bar \phi} + {i \over 2 \sqrt{6}} a \pi_{a} \bar \psi_{A} $$
$$ - \sqrt{3 \over 2} \sigma^{2}a^{2} \bar \psi_{A} + {5i \over 4 \sqrt{2}}
\phi \bar \rchi_{B} \rchi^{B} \bar \rchi_{A} $$
$$-{1 \over 8 \sqrt{6}} \bar \psi^{B} \psi_{A} \bar \psi_{B} - {i \over 4
\sqrt{2}} \phi \psi_{B} \bar \psi^{B} \bar \rchi_{A} $$
$$ +{5 \over 4 \sqrt{6}} \rchi^{B}  \bar \psi_{B} \bar \rchi_{A} - {\sqrt{3}
\over 4 \sqrt{2}} \bar \psi_{B} \rchi_{A} \bar \rchi^{B} $$
$$ - {1 \over 2 \sqrt{6}} \rchi^{B} \bar \rchi_{B}  \bar \psi_{A} \eqno(5.12)$$
Quantum mechanically, one replaces the Dirac brackets by the anti-commutators
if both arguments are odd or commutators if
otherwise:

$$ [E_{1} , E_{2}] = i [E_{1} , E_{2}]^{*} ~,~ [O , E] = i [O , E]^{*} ~,~
\{O_{1} , O_{2}\} = i [O_{1} , O_{2}]^{*}_{+} ~.\eqno(5.13) $$
Here, we use the convention $\hbar = 1 $. And the only non-zero
(anti-)commutators relations are:
$$ \{\rchi_{A} , \bar \rchi_{A} \} = \epsilon_{AB}~, ~
\{\psi_{A} , \bar \psi_{A}\}  = - \epsilon_{AB}~, ~
 [a , \pi_{a}] = i~, ~ [\phi , \pi_{\phi}] = i~,
{}~ [\bar \phi, \pi_{\bar \phi}] = i ~.\eqno(5.14) $$
Here we choose $ (\rchi_{A} , \psi_{A} , a , \phi , \bar \phi) $ to be the
coordinates of the configuration space, and
$ \bar \rchi_{A} , \bar \psi_{A} , \pi_{a} , \pi_{\phi} , \pi_{\bar \phi} $ to
be the momentum operators in this representation.
Hence

$$ \bar \rchi_{A} \rightarrow {\pt \over \pt \rchi_{A}} ,~ \bar \psi_{A}
\rightarrow {\pt \over \pt \psi_{A}},~
 \pi_{a} \rightarrow {\pt \over \pt a} , ~\pi_{\phi} \rightarrow -i {\pt \over
\pt \phi},~
 \pi_{\bar \phi} \rightarrow -i {\pt \over \pt \phi}  \eqno(5.15)$$
Following the ordering used in ref.[5],  we put all the fermionic derivatives
in  $S_{A}$ on the right.
 In $ \bar S_{A} $, all the fermonic
derivatives are on the left. And the supersymmetry generators have the operator
form

$$ S_{A} = -{i \over \sqrt{2}} (1 + \phi \bar \phi) \rchi_{A} {\pt \over  \pt
\phi}
 - {1 \over 2 \sqrt{6}} a \psi_{A} {\pt \over \pt a} $$
$$ - \sqrt{3 \over 2} \sigma^{2}a^{2} \psi_{A}
- {5i \over 4 \sqrt{2}} \bar \phi \rchi_{A} \rchi^{B} {\pt \over \pt
\rchi^{B}}$$
$$-{1 \over 8 \sqrt{6}} \psi_{B} \psi^{B} {\pt \over \pt \psi^{A}}
- {i \over 4 \sqrt{2}} \bar \phi \rchi_{A} \psi^{B} {\pt \over \pt \psi^{B}} $$
$$-{5 \over 4 \sqrt{6}} \rchi_{A} \psi^{B}{\pt \over \pt \rchi^{B}}
 + {\sqrt{3} \over 4 \sqrt{2}} \rchi^{B} \psi_{B} {\pt \over \pt \rchi^{A}} $$
$$+ {1 \over 2 \sqrt{6}} \psi_{A} \rchi^{B} {\pt \over \pt \rchi^{B}}
\eqno(5.16a)$$
and


$$ \bar S_{A} = {i \over \sqrt{2}} (1 + \phi \bar \phi) {\pt \over \pt
\rchi^{A}}
 {\pt \over \pt \bar \phi} + {1 \over 2 \sqrt{6}} a {\pt \over \pt a} {\pt
\over \pt \psi^{A}} $$
$$ - \sqrt{3 \over 2} \sigma^{2}a^{2} {\pt \over  \pt \psi^{A}}
+ {5i \over 4 \sqrt{2}} \phi {\pt \over \pt \rchi^{A}} {\pt \over \rchi^{B}}
\rchi^{B} $$
$$-{1 \over 8 \sqrt{6}} \epsilon^{BC} { \pt \over \pt \psi^{B}} {\pt \over \pt
\rchi^{C}} \psi_{A}
 - {i \over 4 \sqrt{2}} \phi {\pt \over \pt \psi^{B}} {\pt \over \pt \rchi^{A}}
\psi^{B} $$
$$ -{5 \over 4 \sqrt{6}} {\pt \over \pt \psi^{B}} {\pt \over \pt \rchi^{A}}
\rchi^{B}
- {\sqrt{3} \over 4 \sqrt{2}} \epsilon^{BC} {\pt\over \pt \psi^{B}} {\pt \over
\pt \rchi^{C}} \rchi_{A}$$
$$ - {1 \over 2 \sqrt{6}} {\pt \over \pt \psi^{A}} {\pt \over \pt \rchi^{B}}
\rchi^{B} \eqno(5.16b)$$

We now proceed to find the wavefunction of our model.
The Lorentz constraint $ J_{AB} $ is easy to solve. It tells us the
wave function should be a  Lorentz scalar. In our model, $ J_{AB} = \psi_{(A}
\bar \psi_{B)} - \rchi_{(A} \bar \rchi_{B)} $.
 We can easily see that the most general form of the wave function which
satisfies the Lorentz constraint is

$$ \Psi = A + iB \psi^{C} \psi_{C} + C \psi^{C} \rchi_{C} + iD \rchi^{C}
\rchi_{C} + E \psi^{C} \psi_{C} \rchi^{C} \rchi_{C}~, \eqno (5.17)$$
where $A$, $B$, $C$, $D$, and $E$ are functions of $a$, $\phi$ and $\bar \phi$
only. The factors of i are chosen to simplify the future results.
The next step is to solve the supersymmetry constraints $ S_{A} \Psi = 0 $ and
$ \bar S_{A'} \Psi = 0 $.
Since the wave function (5.17) is of  even order in fermonic variables and
stops at order four, the
 $S_{A} \Psi = 0$ and $\bar S_{A} \Psi = 0$ will be of
   odd order in fermonic variables and stop at order three.
 Hence we will get four equations from  $ S_{A} \Psi = 0 $ and  another four
equations from $ \bar S_{A'} \Psi = 0 $.

$$ -{ i \over \sqrt{2}} (1 + \phi \bar \phi) {\pt A \over \pt \phi} = 0~,
\eqno(5.18a)$$
$$ -{ a \over 2 \sqrt{6}} {\pt A \over \pt a} - \sqrt{3 \over 2} \sigma^{2}
a^{2} A = 0~,  \eqno(5.18b)$$
$$ (1 + \phi \bar \phi) { \pt B \over \pt \phi} + {1 \over 2} \bar \phi B
+ { a \over 4 \sqrt{3}} {\pt C \over \pt a} - {7 \over 4 \sqrt{3}} C +
{\sqrt{3} \over 2}  \sigma^{2} a^{2} C = 0~, \eqno(5.18c) $$
$$ {a \over \sqrt{3}} {\pt D \over \pt a} + 2 \sqrt{3} \sigma^{2} a^{2} D -
\sqrt{3} D
- (1 + \phi \bar \phi) { C \over \phi} - {3 \over 2} \bar \phi C = 0~,
\eqno(5.18d) $$
$$ i \sqrt{2} (1 + \phi \bar \phi) {\pt E \over \pt \bar \phi} = 0~,
\eqno(5.19a)$$
$$ {a \over \sqrt{6}} { \pt E \over \pt a} - \sqrt{6} \sigma^{2} a^{2} E = 0~,
\eqno(5.19b) $$
$$ {a \over \sqrt{3}} {\pt B \over \pt a} - 2 \sqrt{3} \sigma^{2} a^{2} B -
\sqrt{3} B
 + (1 + \phi \bar \phi) {\pt C \over \pt \bar \phi} + {3 \over 2} \phi C = 0
{}~,\eqno(5.19c)$$
$$  (1 + \phi \bar \phi) {\pt D \over \pt \bar \phi} + {1 \over 2} \phi D
- {a \over 4 \sqrt{3}} {\pt C \over \pt a} + {7 \over 4 \sqrt{3}} C + {\sqrt{3}
\over 2} \sigma^{2} a^{2} C = 0 ~. \eqno(5.19d)$$
We can see that (5.18a), (5.18b) and (5.19a), (5.19b)
constitute  decoupled equations for $A$ and $E$, respectively.
 They have the general solution.
$$ A = f(\bar \phi) \exp({-3 \sigma^{2} a^{2}})~,~
 E = g(\phi) \exp ({3 \sigma^{2} a^{2}}) \eqno(5.20) $$
where $ f , g $ are arbitrary anti-holomorphic and holomorphic functions of
$\phi$, respectively.
 Eq. (5.18c) and (5.18d) are coupled equations between $B$ and $C$ and eq.
(5.19c) and (5.19d) are coupled equations between $C$ and $D$.
 The first step to decouple these equations is as follows.
 Let $ B = \tilde B (1 + \phi \bar \phi)^{- {1 \over 2}} $ ,
$ C = {\tilde C  \over \sqrt{3}}(1 + \phi \bar \phi)^{- {3 \over 2}} $ , $ D =
\tilde D (1 + \phi \bar \phi)^{- {1 \over 2}}$.
 Equations (5.18c), (5.18d), (5.19c) and (5.19d) then  become

$$ (1 + \phi \bar \phi)^{2} {\pt \tilde B \over \pt \phi} + {a \over 12} {\pt
\tilde C \over \pt a}
- {7 \over 12} \tilde C + {1 \over 2} \sigma^{2} a^{2}  \tilde C =
0~,\eqno(5.21a) $$
$$ (1 + \phi \bar \phi)^{2} {\pt \tilde D \over \pt \bar \phi} - {a \over 12}
{\pt \tilde C  \over \pt a}
+ {7 \over 12} \tilde C + {1 \over 2} \sigma^{2} a^{2}  \tilde C = 0~,
\eqno(5.21b) $$
$$ {\pt \tilde C \over \pt \phi} - a {\pt \tilde D \over \pt a} - 6 \sigma^{2}
a^{2} \tilde D + 3 \tilde D = 0 ~,\eqno(5.21c) $$
$$ {\pt \tilde C \over  \pt \bar \phi} + a {\pt \tilde B  \over \pt a} - 6
\sigma^{2} a^{2} \tilde B - 3 \tilde B = 0~. \eqno(5.21d)$$
 From (5.21a) and (5.21d), we can eliminate $\tilde B$ to get a partial
differential equation for $\tilde C$:
$$  (1 + \phi \bar \phi)^{2} {\pt \tilde C \over \pt \bar \phi \pt \phi}
 - {a \over 12} {\pt \over \pt a} \left(a {\pt \tilde C \over \pt a}\right)
+ {5 \over 6} a {\pt \tilde C \over \pt a} + \left[ 3 \sigma^{4} a^{4} + 3
\sigma^{2} a^{2} - {7 \over 4} \right] \tilde C = 0~, \eqno(5.22) $$
and from (5.21b) and (5.21c), we will get another partial differential equation
for $\tilde C$:
$$  (1 + \phi \bar \phi)^{2} {\pt \tilde C \over \pt \bar \phi \pt \phi}
 - {a \over 12} {\pt \over \pt a} \left(a {\pt \tilde C \over \pt a}\right) +
{5 \over 6} a {\pt \tilde C \over \pt a}
 + \left[ 3 \sigma^{4} a^{4} - 3 \sigma^{2} a^{2} - {7 \over 4} \right] \tilde
C = 0~.
 \eqno(5.23) $$
We can see immediately that  $\tilde C = 0$ because the coefficients of
$\sigma^{2} a^{2} \tilde C$ are different for these two equations.
 Using this result, we find
$$ B = h(\bar \phi) (1 + \phi \bar \phi)^{- {1 \over 2}} a^{3}
\exp({3 \sigma^{2} a^{2}})~,~
 C = 0 ~,~ D = k(\phi) (1 + \phi \bar \phi)^{- {1 \over 2}} a^{3} \exp ({-3
\sigma^{2} a^{2}} ) ~. \eqno(5.24)$$

This results can be strengthened as
we will   show that
$\tilde C = 0$ is not a result of the particular ordering used in the above
calculations.
 In fact, we can try the ordering presented in ref. [6]   such that $S_{A}$ and
$\bar S_{A'}$
are hermitian adjoints in the standard inner product, appropiate to the
holomorphic
representation
being used here for the fermions. If one allows for the factor ordering
ambiguity in $S_{A}$
due to the terms cubic in fermions, and insists that $\bar S_{A'}$ be the
hermitian adjoint of $S_{A}$, the operators have the form

$$ S_{A_{new}} = S_{A} + \lambda \psi_{A} + i \mu \bar \phi \rchi_{A}~, $$
$$ \bar S_{A_{new}} = \bar S_{A} + i \left( {7 \over 4 \sqrt{2}} - \mu \right)
\bar \phi \rchi_{A} +
\left({5 \over 4 \sqrt{6}} - \lambda \right) \bar \psi_{A}~, \eqno(5.25) $$
where $S_{A}$ , $\bar S_{A}$ are the ordering used above. Proceeding
 to solve these constraints, we will find another eight equations.
We are only interested in the four equations between $B$, $C$ and $D$ :

$$ (1 + \phi \bar \phi) {\pt B \over \pt \phi} + \left({1 \over 2} - \sqrt{2}
\mu \right) \bar \phi B
+ { a \over 4 \sqrt{3}} {\pt C \over \pt a} - {7 \over 4 \sqrt{3}} C  -
{\lambda \over \sqrt{2}} C
+ {\sqrt{3} \over 2}  \sigma^{2} a^{2} C = 0~, \eqno(5.26a) $$
$$ {a \over \sqrt{3}} {\pt D \over \pt a} + 2 \sqrt{3} \sigma^{2} a^{2} D -
(\sqrt{3}
+ 2 \sqrt{2} \lambda) D - (1 + \phi \bar \phi) {\pt C \over \pt \phi} - {3
\over 2} \bar \phi C + \sqrt{2} \mu \bar \phi C = 0~, \eqno(5.26b) $$
$$ {a \over \sqrt{3}} {\pt B \over \pt a} - 2 \sqrt{3} \sigma^{2} a^{2} B -
\sqrt{3} B
+ \sqrt{2} \left({5\over 2 \sqrt{6}} - 2 \lambda \right)B  $$
$$ + (1 + \phi \bar \phi) {\pt C \over \pt \bar \phi}
+ {3 \over 2} \phi C  -\sqrt{2} \left({7 \over 4 \sqrt{2}} - \mu\right) \phi C=
0 ~, \eqno(5.26c)$$
$$  (1 + \phi \bar \phi) {\pt D \over \pt \bar \phi} + {1 \over 2} \phi D
 -{1 \over \sqrt{2}} \left({7 \over 2 \sqrt{2}} -2 \mu \right) \phi D - {a
\over 4 \sqrt{3}} {{\ pt C} \over  {\pt a}}
 $$
$$ + {7 \over 4 \sqrt{3}} C -{1 \over \sqrt{2}} \left({5 \over 4 \sqrt{6}}
-\lambda \right) C + {\sqrt{3} \over 2} \sigma^{2} a^{2} C = 0 ~.\eqno(5.26d)
$$
We must set $\mu = {7 \over 8 \sqrt{2}}$ in order to get rid of $\bar \phi C$
in (5.26b) and (5.26c) consistently,
and the only freedom left to get consistent equations for $\tilde C$ is from
$\lambda$.
By setting $ B = \tilde B (1 + \phi \bar \phi)^{3 \over 8} $,
$ C = {\tilde C  \over \sqrt{3}}(1 + \phi \bar \phi)^{- {5 \over 8}} $,
$ D = \tilde D (1 + \phi \bar \phi)^{3 \over 8}$. we can again get two
decoupled equations for $\tilde C$ from the above 4 equations.
 Again the coeffecient of $\sigma^{2}a^{2} \tilde C$ for one equation
is $ -{7 \over 4} $ and the coefficient of  $\sigma^{2}a^{2} \tilde C$ for the
other
equation is ${17 \over 4}$.
 Hence,  we are led  again to  $\tilde C = 0$, showing that the two most
 interesting orderings give  $\tilde C = 0$.

\medbreak
\noindent
{\bf VI. Discussion and Conclusion.}

We would like to compare our results to the ones in ref. [5]. There,
 $\rrrm^2 $ was used  as  K\"ahler manifold and a different result for
 $\tilde C$ was obtained. We would like to  investigate  here
whether we can get
a similar result for $\tilde C$ in  our model in the case of the
$\rrrm^{2}$ K\"ahler manifold.
The K\"ahler potential  would be just  $\phi \bar \phi$,
the K\"ahler metric is $g_{\phi \bar \phi} = 1$ and the Levi-Cita connections
are zero.
Repeating the steps described  in sections II--V, we find out that
 the structure of the supersymmetry constraints are the same for these two
K\"ahler manifolds.
The reason is that the K\"ahler metric and the connection only enter the
Lagrangian
 through
the spin-${1 \over 2}$ field $\rchi_{A}$ and no other terms.
So, there is only a change in
the
 coefficient of $\bar \phi \rchi_{A} \rchi^{B}
{\pt  \over \pt \rchi^{B}}$ in $S_{A}$ and
 the corresponding term in $\bar S_{A}$, the rest being equivalent to
 put $\phi \bar \phi = 0$
in the necessary coefficients. The supersymmetry constraints are then

$$ S_{A} = -2 \sqrt{3} i \rchi_{A} { \pt \over \pt \phi} - a \psi_{A} {\pt
\over \pt a}
 - 6 \sigma^{2}a^{2} \psi_{A} - {i \sqrt{3} \over 2} \bar \phi \rchi_{A}
\rchi^{B} {\pt \over \pt \rchi^{B}} $$
$$-{1\over 4} \psi_{B} \psi^{B} {\pt \over \pt \psi^{A}} - {i \sqrt{3} \over 2}
\bar \phi \rchi_{A} \psi^{B} {\pt \over \pt \psi^{B}}
 -{5 \over 2} \rchi_{A} \psi^{B}{\pt \over \pt \rchi^{B}} + {3 \over 2}
\rchi^{B} \psi_{B} {\pt \over \pt \rchi^{A}}
 + \psi_{A} \rchi^{B} { \pt \over \pt \rchi^{B}} \eqno(6.1) $$
and

$$ \bar S_{A} = 2 \sqrt{3} i {\pt \over \pt \rchi^{A}} {\pt \over \pt \bar
\phi} + a {\pt \over \pt a} {\pt \over \pt \psi^{A}}
- 6 \sigma^{2}a^{2} {\pt \over \pt \psi^{A}} + {i \sqrt{3} \over 2} \phi {\pt
\over \pt  \rchi^{A}} {\pt \over \pt \rchi^{B}} \rchi^{B}
-{1 \over 4} \epsilon^{BC} {\pt\over \pt \psi^{B}} {\pt \over \pt \rchi^{C}}
\psi_{A} $$
 $$- {i \sqrt{3} \over 2} \phi {\pt \over \pt \psi^{B}} {\pt \over \pt
\rchi^{A}} \psi^{B}
 - {5 \over 2} {\pt \over \pt \psi^{B}} {\pt \over \pt \rchi^{A}} \rchi^{B}
 - {3 \over 2} \epsilon^{BC} {\pt \over \pt \psi^{B}} {\pt \over  \pt
\rchi^{C}} \rchi_{A}
 - {\pt \over \pt \psi^{A}} {\pt \over \pt \rchi^{B}} \rchi^{B}~. \eqno(6.2)$$
Solving for the $S_{A} \Psi = 0$ and $\bar S_{A} \Psi = 0$,we obtain eight
equations
where the   four equations between $B$ , $C$ and $D$ are:

$$ { \pt B \over \pt \phi} + {1 \over 2} \bar \phi B + { a \over 4 \sqrt{3}}
{\pt C \over \pt a}
 - {7 \over 4 \sqrt{3}} C + {\sqrt{3} \over 2}  \sigma^{2} a^{2} C = 0
{}~,\eqno(6.3a) $$
$$ {a \over \sqrt{3}} {\pt D \over \pt  a} + 2 \sqrt{3} \sigma^{2} a^{2} D
- \sqrt{3} D - {\pt C \over \pt \phi} - {1 \over 2} \bar \phi C  = 0
{}~,\eqno(6.3b)$$
$$ {a \over \sqrt{3}} {\pt  B \over \pt  a} - 2 \sqrt{3} \sigma^{2} a^{2} B
- \sqrt{3} B + {\pt C \over \pt \bar \phi} + {1 \over 2} \phi C  = 0~,
\eqno(6.3c) $$
$$ {\pt D \over \pt \bar \phi} + {1 \over 2} \phi D - {a \over 4 \sqrt{3}} {\pt
C \over \pt a}
 + {7 \over 4 \sqrt{3}} C + {\sqrt{3} \over 2} \sigma^{2} a^{2} C = 0
{}~.\eqno(6.3d)$$
Making the   substitution $B = \tilde B \exp ({-{1 \over 2} \phi \bar \phi})$ ,
 $C ={\tilde C \over \sqrt{3}} \exp ({-{1 \over 2} \phi \bar \phi})$
 and $D = \tilde D \exp ({-{1 \over 2} \phi \bar \phi})$ [5], the above four
equations become

$$ {\pt \tilde B \over \pt \phi} + {a \over 12} {\pt \tilde C \over \pt a}
- {7 \over 12} \tilde C + {1 \over 2} \sigma^{2} a^{2}  \tilde C = 0
{}~,\eqno(6.4a) $$
$$ {\pt \tilde D \over \pt \bar \phi} - {a \over 12} {\pt \tilde C \over \pt a}
+ {7 \over 12} \tilde C + {1 \over 2} \sigma^{2} a^{2}  \tilde C = 0
{}~,\eqno(6.4b) $$
$$ {\pt \tilde C \over \pt \phi} - a {\pt \tilde D \over \pt a}
- 6 \sigma^{2} a^{2} \tilde D + 3 \tilde D = 0 ~ \eqno(6.4c)$$
$$ {\pt \tilde C \over \pt \bar \phi} + a {\pt \tilde B \over \pt a} - 6
\sigma^{2} a^{2} \tilde B - 3 \tilde B = 0 ~.\eqno(6.4d) $$
This set of equations are exactly the same as (5.21a -- d)
 if we put $\phi \bar \phi = 0$ in there.
So,  $\tilde C = 0$ does not depend on the value of $\phi \bar \phi$.
 We conclude, therefore, that for $\rrrm^{2}$ as the K\"ahler manifold, $\tilde
C =0$.
These results seem to suggest that whatever  K\"ahler manifold one uses,
 we reach the same conclusion. The reason for the apparent
differences with respect to ref.[5] may lie in the fact that the model
used in ref. [5] was derived from ref. [28], while ours comes directly from
ref. [27].

Summarizing our work, in section II the more general theory of $ N = 1 $
 supergravity with supermatter was studied using a
canonical approach. The supersymmetry and gauge constraint generators were also
found.
In section III we described the Ans\"atze for the the gravitational field,
 spin-${3 \over 2}$ field and the gauge vector field $V^{a}_{\mu}$ as well as
the scalar
 fields and the  corresponding fermionic partners. In section IV,
after a dimensional reduction, we derived the supersymmetric constraints
for our one-dimensional model, where one has assumed a FRW closed geometry.
In section 5, we solved the supersymmetry constraints for the case of a
 $S^{2}$  K\"ahler manifold, taking   the spin-1 field to be zero as well
as its fermionic partner.
 We found that one of the middle states is missing as  $C =0$, and the other
middle states have the
 simple form $B = g(\bar \phi) (1 + \phi \bar \phi)^{-{1 \over2}} a^{3} \exp
({3 \sigma^{2} a^{2}})$ and
 $D = f(\phi) (1 + \phi \bar \phi)^{-{1 \over2}} a^{3}  \exp({-3 \sigma^{2}
a^{2}})$.
However, these are not  wormhole states. From [6], the wormhole  wavefunction
has
 the form   ${\rm prefactor}\times\exp ({-(3 \sigma^{2}
a^{2} + 3 \sigma^{2} a^{2} \cosh(\rho))})$,
where $\phi = \rho \exp (i\theta) $, and such behaviour is not
provided by $B$ and $D$.
In order to investigate this issue we repeated our study but for a $\rrrm^2$
 K\"ahler manifold. Once again, we   have  $C = 0$.
Thus, it seems that the results obtained from the framework presented in ref.
[27]
are quite general.

The bosonic and filled fermion states have  the simple form
 $A= f(\bar \phi) \exp ({-3 \sigma^{2} a^{2}})$ and
 $ E =  g(\phi) \exp ({3 \sigma^{2} a^{2}})$, respectively.
This corresponds to  the Hartle-Hawking state.
It is very puzzling that the wormhole state is missing.
 A similar problem occured in ref. [3]. There,  the only bosonic physical state
is the wormhole solution
 but the Hartle-Hawking state was missing.
However,  if we use a different definition of homogeneity [23], we will get
 the Hartle-Hawking state as the bosonic state but then the wormhole state is
 missing. We suspect that similar behaviour  is  occurring here.

In the future we will extend the framework presented in this paper in
two directions of study: the inclusion of all supermatter fields in the process
of solving the supersymmetry constraints [37],  and generalizing
the work  the case of
 a Bianchi-IX universe.
 It will be interesting to see if the same type
of results  occur there.

\noindent
{\bf ACKNOWLEDGEMENTS}

A.D.Y.C. thanks the Croucher Foundation of Hong Kong for financial support.
\break
P.R.L.V.M. is very grateful to Prof. V.N. Melnikov and to the Organizing
Committee of the International School-Seminar
"Multidimensional Gravity and Cosmology"  for all their
kind assistance.
P.R.L.V.M. also gratefully acknowledges the support of
a Human Capital and Mobility
grant from the European Union (Program ERB4001GT930714).

\vskip .2 true in
\noindent
{\bf REFERENCES}

{\rm

\advance\leftskip by 4em
\parindent = -4em

[1] P.D. D'Eath, Phys.~Rev.~D {\bf 29}, 2199 (1984).

[2] P.D. D'Eath, S.W. Hawking and O. Obreg\'on, Phys.~Lett.~{\bf 300}B, 44
(1993).

[3] P.D. D'Eath, Phys.~Rev.~D {\bf 48}, 713 (1993).

[4] P.D. D'Eath and D.I. Hughes, Phys.~Lett.~{\bf 214}B, 498 (1988).

[5] P.D. D'Eath and D.I. Hughes, Nucl.~Phys.~B {\bf 378}, 381 (1992).

[6] L.J. Alty, P.D. D'Eath and H.F. Dowker, Phys.~Rev.~D {\bf 46}, 4402
(1992).

[7] D.I. Hughes, Ph.D.~thesis, University of Cambridge (1990), unpublished.

[8] P.D. D'Eath, Phys. Lett. B{\bf 320}, 20 (1994).

[9]  A.D.Y. Cheng, P.D. D'Eath and
P.R.L.V. Moniz, Phys. Rev. D{\bf 49} (1994) 5246.

[10]  A.D.Y. Cheng, P.D. D'Eath and
P.R.L.V. Moniz, DAMTP R94/20, {\it Quantization of Bianchi Models in N=1 Sugra
with a
Cosmological Constant.} gr-qc 9406047

[11] P.D. D'Eath,  Phys.~Lett.~B {\bf 321}, 368 (1994).

[12]  P.D. D'Eath,  DAMTP report.

[13] A.D.Y. Cheng, P.D. D'Eath and
P.R.L.V. Moniz, {\it Canonical Formulation  of N=1
Supergravity with Supermatter}~DAMTP R94/13, submitted to
Physical Review D.

[14] R. Graham, Phys.~Rev.~Lett.~{\bf 67}, 1381 (1991).

[15] R. Graham, Phys. Rev. D{\bf 48}, 1602 (1993).

[16] R. Graham and J. Bene,  Phys. Rev. D{\bf 49}, 799 (1994) and
references therein.

[17] T. Jacobson, Class. Quantum Grav. {\bf 5}, 923, (1988).

[18]  T. Sano and J. Shiraishi, Nucl. Phys. B{\bf 410}, 423, (1993).

[19] R. Capovilla and O. Obregon, CIEA-GR-9402, gr-qc:9402043.

[20] H.-J. Matschull, DESY 94-037, gr-qc:9403034.

[21] P. D. D'Eath and P. V. Moniz, {\it  Supersymmetric Quantum  Cosmology},
work in preparation.

[22] C. Teitelboim, Phys.~Rev.~Lett.~{\bf 38}, 1106 (1977).

[23] R. Graham and H. Luckock, Sidney Univ. Maths. Report. 93-06,
gr-qc 9311004.

[24] S.W. Hawking and D.N. Page, Phys.~Rev.~D {\bf 42}, 2655 (1990).

[25] J.B. Hartle and S.W. Hawking, Phys.~Rev.~D {\bf 28}, 2960 (1983).

[26] M. Asano, M. Tanimoto and N. Yoshino, Phys.~Lett.~{\bf 314}B, 303 (1993).

[27]  J. Wess and J. Bagger, {\it Supersymmetry and Supergravity},
2nd.~ed. (Princeton University Press, 1992).

[28] A. Das. M. Fishler and M. Rocek,  Phys.~Lett.~B {\bf 69}, 186 (1977).

[29] P.V. Moniz and J. Mour\~ao, Class.  Quantum Grav. {\bf 8}, (1991) 1815 and
references therein.

[30]  M.C. Bento, O. Bertolami, J. Mour\~ao,  P.V. Moniz and
P. S\'a, Class.  Quantum Grav. {\bf 10} (1993) 285

[31] P.R.L.V. Moniz, Ph.D.~thesis, University of Lisbon (1993), unpublished.

[32] J.E. Nelson and C. Teitelboim, Ann.~Phys. (N.Y.) {\bf 116} (1978)
86.

[33] P.D. D'Eath and J.J. Halliwell, Phys.~Rev.~D {\bf 35} (1987) 1100.

[34] P. van Nieuwenhuizen, Phys.~Rep. {\bf 68} (1981) 189.

[35]  L.D. Faddeev and A.A. Slavnov, {\it Gauge Fields} (Benjamin,
Reading, Mass., 1980).

[36] E. Witten and J. Bagger, Phys.~Lett.~B {\bf 115} (1982).

[37] A.D.Y. Cheng, P.D. D'Eath and
P.R.L.V. Moniz, work in preparation

\ \ \ }

\bye